\documentclass[fleqn]{elsart}

\usepackage[dvips]{graphicx}           % version 0.7h
\usepackage{amsmath}
\usepackage{amssymb}

\usepackage[ansinew]{inputenc}

\newcommand{\MB}{\mbox{MgB$_2$}}
\newcommand{\Tc}{\ensuremath{T_c}}

\newcommand{\omlog}{\ensuremath{\langle\omega_{\rm log} \rangle}}
\newcommand{\omsquare}{\ensuremath{\langle\omega^2 \rangle}}
\newcommand{\WN}{\mbox{cm$^{-1}$}}

\begin{document}

% ----------------------------------------------------------------

\begin{frontmatter} % Elsevier
%\volume{96}
%\issue{4}
%\firstpage{183}
%\lastpage{187}
%\nocopyright
\title{Calculated elastic and electronic properties of MgB$_\mathbf{2}$ at high pressures}

\author{I. Loa and K. Syassen}
\address{Max-Planck-Institut f\"{u}r Festk\"{o}rperforschung,
            Heisenbergstr.\ 1,
            D-70569 Stuttgart,
            Germany}

\received{February 26, 2001}
\accepted{March 2, 2001 by M. Cardona\\ Solid State Commun., in press}
%\communicated{M. Cardona}
%\acceptedprefix{; accepted }
%\communicatedprefix{ by }

\begin{keyword}
A. Superconductors; D. Crystal binding and equation of state; D. Electronic band
structure; A. Metals\\
   \hspace*{\fill}
\end{keyword}

\begin{abstract}
The effect of high pressure on structural and electronic properties of the novel
superconductor \MB\ has been calculated using the full-potential linearized
augmented-plane-wave method. Despite the layered crystal structure of \MB\ nearly
isotropic compression (bulk modulus $B_0=140.1(6)$~GPa) is found with only a
1.2\% decrease of the $c/a$ ratio at 10~GPa. The effect of pressure on the
critical temperature has been estimated on the basis of BCS theory and good
agreement with experimental data is found. Our results suggest that it is a
combination of increasing phonon frequency and decreasing electronic density of
states at the Fermi level which leads to the observed decrease of the critical
temperature under pressure.
\end{abstract}

\end{frontmatter}

% ----------------------------------------------------------------
A recent report of Nagamatsu \textit{et al.}\/ \cite{NNMZ01} on superconductivity
in
\MB\ with a relatively high critical temperature of $\Tc = 39$~K has raised great
interest in this compound. At ambient conditions \MB\ has a layered hexagonal
crystal structure of the AlB$_2$ type (space group
\textit{P\,6/m\,m\,m}, $Z=1$) \cite{JM54} and shows metallic conductivity
\cite{FOBL01p,JPKK01p}.
\textit{Ab initio} calculations of the electronic properties
\cite{BSA01p,KMBA01p,IM00} indicate the coexistence of strong covalent bonding
within the honeycomb B layers and metallic bonding between the Mg and B layers.
On the basis of BCS theory, the high critical temperature was attributed to a
fortunate combination of strong bonding, `reasonable' electronic density of
states at the Fermi level, and high phonon frequencies \cite{KMBA01p}. This view
of phonon-mediated superconductivity is supported by the observation of a sizable
boron isotope effect with $\Delta \Tc =+1.0$~K for substitution of $^{11}$B by
$^{10}$B \cite{BLPC01}. The effect of high pressure on \Tc\ was studied by Lorenz
\textit{et al.}\/ \cite{LMC01p} and Monteverde \textit{et al.}\/ \cite{MNRR01p}.
Both groups observed a decrease of \Tc\ with increasing pressure. Hirsch, on the
other hand, predicted on the basis of his theory of `hole superconductivity' an
increase of \Tc\ provided that the intralayer B--B distances decrease and that no
charge transfer between the Mg and B layers occurs \cite{Hir01p}.

Motivated by the experimental high-pressure study of \MB\ and failure of the
theoretical prediction  we have calculated the effect of high pressure on
structural and electronic properties of \MB\ using the full-potential linearized
augmented-plane-wave method (FPLAPW). Total-energy calculations as a function of
unit cell volume with optimized $c/a$ lattice parameter ratios yield the equation
of state. Despite the layered crystal structure we find nearly isotropic
compression. We furthermore estimated the pressure dependence of the transition
temperature
\Tc. Based on BCS theory with some simplifying assumptions we find good agreement
with the experiment. Our results suggest that it is a combination of increasing
phonon frequency and decreasing electronic density of states at the Fermi level
which leads to the observed decrease of the critical temperature under pressure.

We have used the first-principles FPLAPW method as implemented in the WIEN97 code
\cite{soft:WIEN97}. For the exchange-correlation potential we employed the
generalized gradient approximation of Ref.~\cite{PBE96}. Scalar-relativistic
corrections were included. Atomic spheres were kept constant in size for the
total energy calculations \cite{param:TE} whereas they were scaled with the unit
cell volume for the DOS calculations \cite{param:DOS}. For $k$-point sampling 296
points were used in the irreducible wedge of the Brillouin zone (4536 in total).
The Mg $2s$ and $2p$ states were treated as band states using the local orbital
extension of the LAPW method \cite{soft:WIEN97,Sin91}.

\begin{figure}[bt]
     \centering
     \includegraphics[scale=0.5,clip]{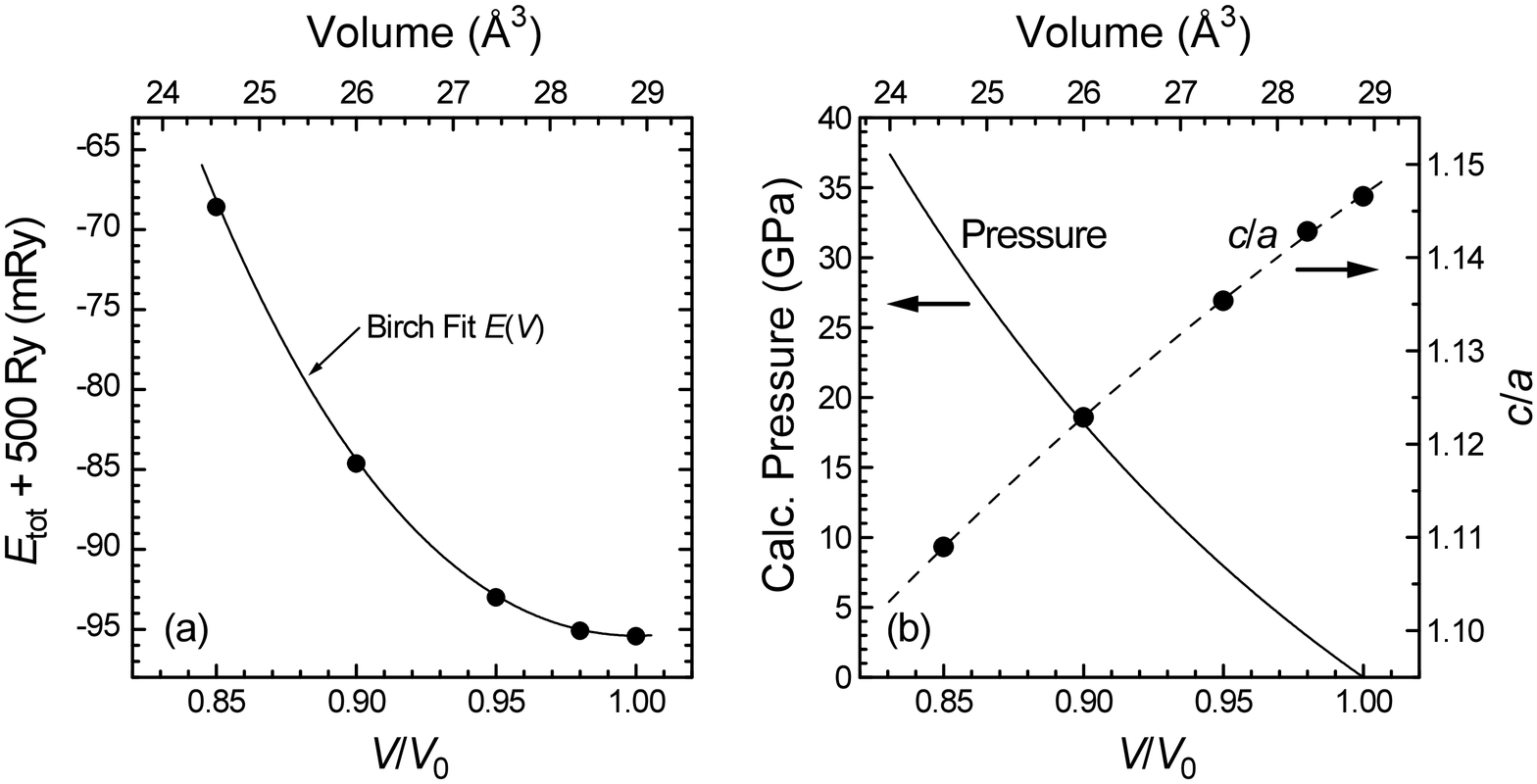}
     \vspace{3mm}
     \caption{(a) Total energy of \MB\ as a function of reduced volume (symbols).
     The solid line represents a fit of a Birch-type $E(V)$ relation to the data.
     (b) Calculated pressure (solid line) and optimized $c/a$ ratio (symbols) as
     a function of reduced volume. The dashed line represents a quadratic fit to
     the $c/a$ data \cite{param:CoAFit}. The upper axes show the unit cell
     volume.}
     \label{fig:structure}
\end{figure}

The total energy of \MB\ was calculated for reduced volumes $V/V_0$ in the range
0.85--1.00 [Fig.~\ref{fig:structure}(a)] (the subscript `0' marks zero-pressure
parameters throughout this paper). At each volume, the $c/a$ lattice parameter
ratio was optimized [Fig.~\ref{fig:structure}(b)]. An energy-vs-volume relation,
obtained from integrating the Birch $P(V)$ function \cite{Bir78}, was fitted to
the total-energy data to obtain the bulk modulus $B_0 = 140.1(6)$~GPa, its
pressure derivative at zero pressure $B' = 3.93(14)$, and the unit cell volume
$V_0 = 28.888(11)$~{\AA}$^3$ (the errors denote the standard error of the
least-squares fit). The corresponding pressure-vs-volume relation is shown in
Fig.~\ref{fig:structure}(b). We did not consider possible structural phase
transitions.

The calculated zero-pressure, zero-temperature $V_0 = 28.888(11)$~{\AA}$^3$ and $c/a
= 1.1468(3)$ are in good agreement with the experimental room-temperature data
$V_0 = 29.00$~{\AA}$^3$ and $c/a = 1.142$ \cite{JM54,LMC01p}. A rather isotropic
compressibility is indicated by the small changes of the $c/a$ ratio. This shows
that the intra- and the interlayer bonding are of similar strength and that \MB,
in this respect, is distinct from layered compounds such as (intercalated)
graphite where the sheets are only weekly bonded. This view is in agreement with
the finding of Belashchenko \textit{et al.}\ that the structure of \MB\ is
characterized by the coexistence of strong covalent bonding within the B layers
and delocalized, `metallic-type' bonding between the Mg and B sheets
\cite{BSA01p}.

\begin{figure}[bt]%
   \centering%
   \begin{minipage}{11cm}%
   \includegraphics[width=11cm,clip]{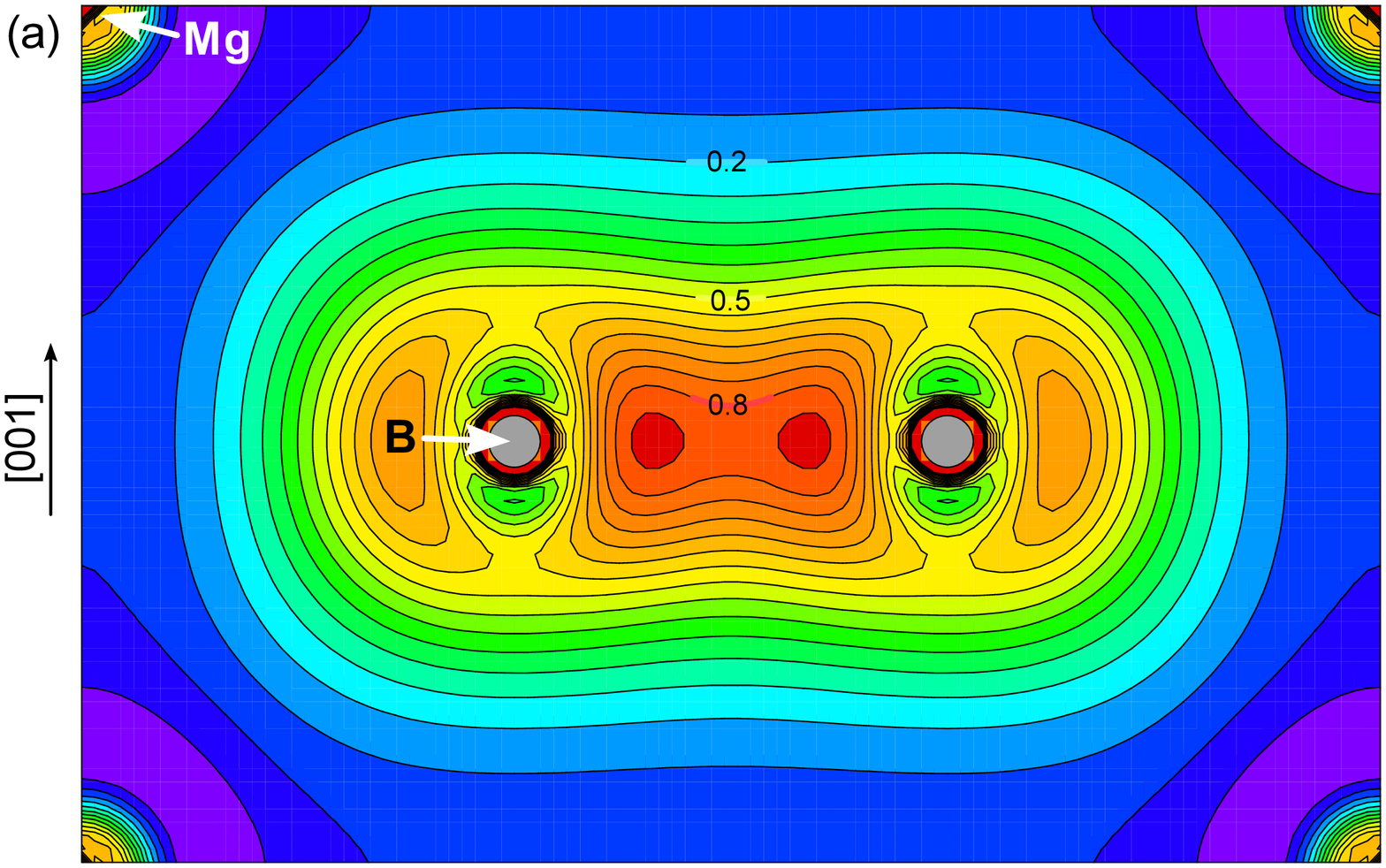}\\[1mm]
   \includegraphics[width=11cm,clip]{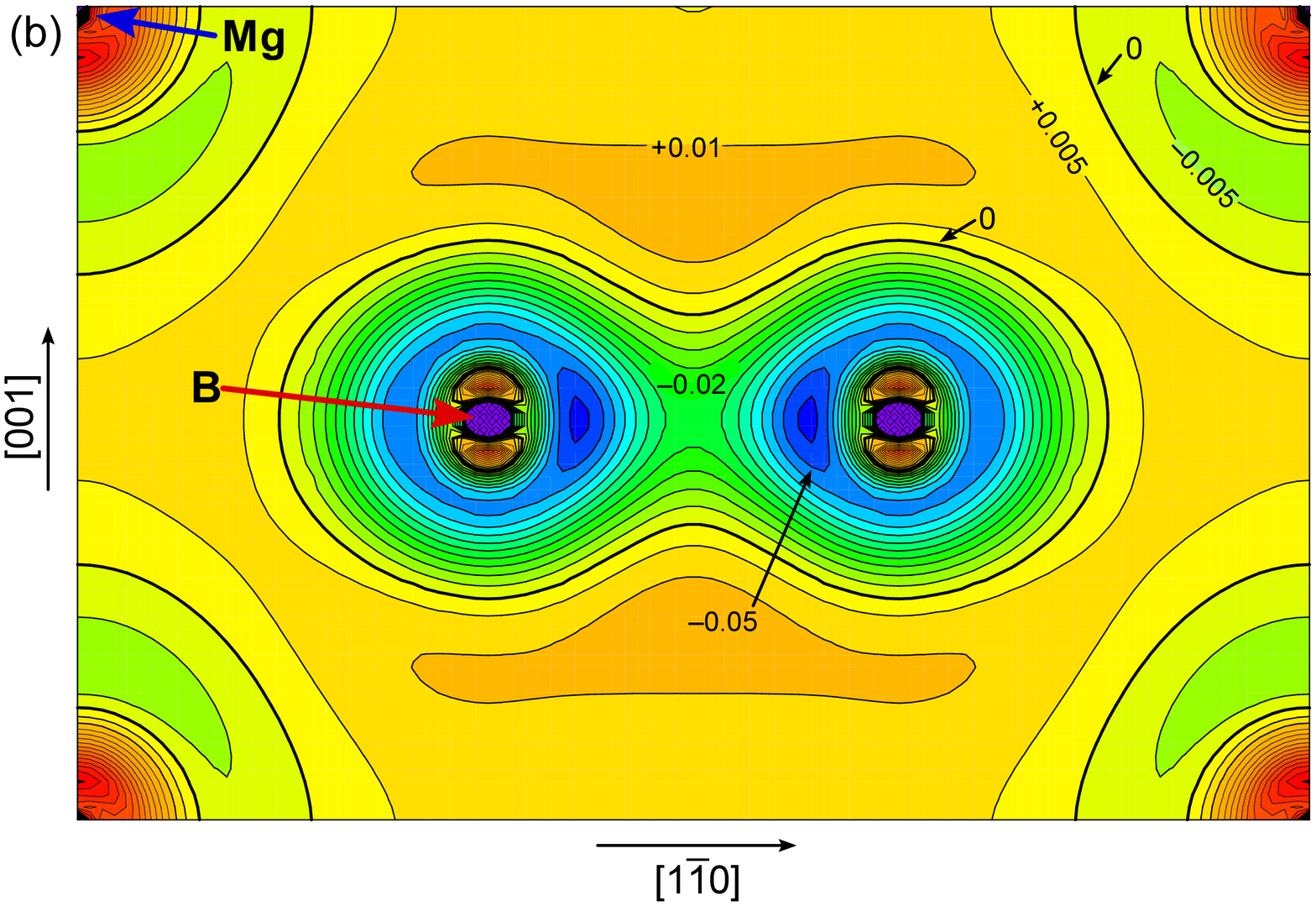}
   \end{minipage}
     \vspace{3mm}
     \caption{(a) Charge density distribution of \MB\ in the (110) plane with
     contour lines at intervals of 0.05~$e$/{\AA}$^3$. (b) Change of the charge
     density distribution $\rho$ at 20~GPa compared to 0~GPa ($\rho_{20}
     V_{20}/V_0 - \rho_0$; the 20-GPa charge density has been scaled by
     $V_{20}/V_0$ to take into account the change of volume). Contour
     lines are at intervals of 0.005~$e$/{\AA}$^3$.}
     \label{fig:CD}
\end{figure}

In the context of bonding properties it is also of interest to examine the
changes of the charge distribution at high pressures. Figure~\ref{fig:CD}(a)
shows the charge density distribution in the (110) plane of \MB. Mg nuclei are
located at the corners of the map and B nuclei at the (1/3, 1/2) and (2/3, 1/2)
positions, all of them in the plane of the figure. One can clearly see the
directional, covalent B--B bonds. In addition there is a significant amount of
charge in the interstitial region giving rise to metallic-type bonding between
the Mg and B sheets. The charge distribution and bonding properties of
\MB\ at ambient pressure are discussed in detail in
Refs.~\cite{BSA01p,KMBA01p,IM00}. Figure~\ref{fig:CD}(b) shows the difference
between the charge densities at 20 and 0 GPa. The 20-GPa charge density has been
scaled by $V/V_0$ in order to take into account the change of volume.  There
occurs a transfer of charge from the region between the B ions into the nearby
interstitial region. This can be regarded as a transfer from $\sigma$ to $\pi$
type bonds which has been termed `hole doping' \cite{AP01p}. This trend is also
found in the decomposition of the charge in the boron atomic sphere. These
changes are, however, rather moderate so that the main characteristics of the
bonding in \MB\ remain unchanged at pressures up to 20~GPa.

In the framework of BCS theory, the transition temperature of a monoatomic
superconductor can be expressed according to the McMillan formula
\cite{McM68,KMBA01p} as
\begin{equation}\label{eq:McMillan}
  \Tc = \frac{\langle \omega_{\rm log} \rangle}{1.20}\,
      \exp{\left[-1.02\frac{1+\lambda}{\lambda-\mu^*-\mu^*\lambda}\right]} \quad.
\end{equation}

Here, $\lambda = N(0)\langle I^2 \rangle/M \langle \omega^2 \rangle$ denotes the
electron--phonon coupling constant, $\mu^*$ the Coulomb pseudopotential, $N(0)$
the electronic density of states at the Fermi level, $\langle I^2 \rangle$ the
averaged electron--ion matrix element, $M$ the atomic mass, $\langle \omega^2
\rangle$ the averaged square of a characteristic phonon frequency, and $\langle
\omega_{\rm log} \rangle$ the logarithmically averaged phonon frequency.

\begin{figure}[bt]
     \centering
     \includegraphics[scale=0.5,clip]{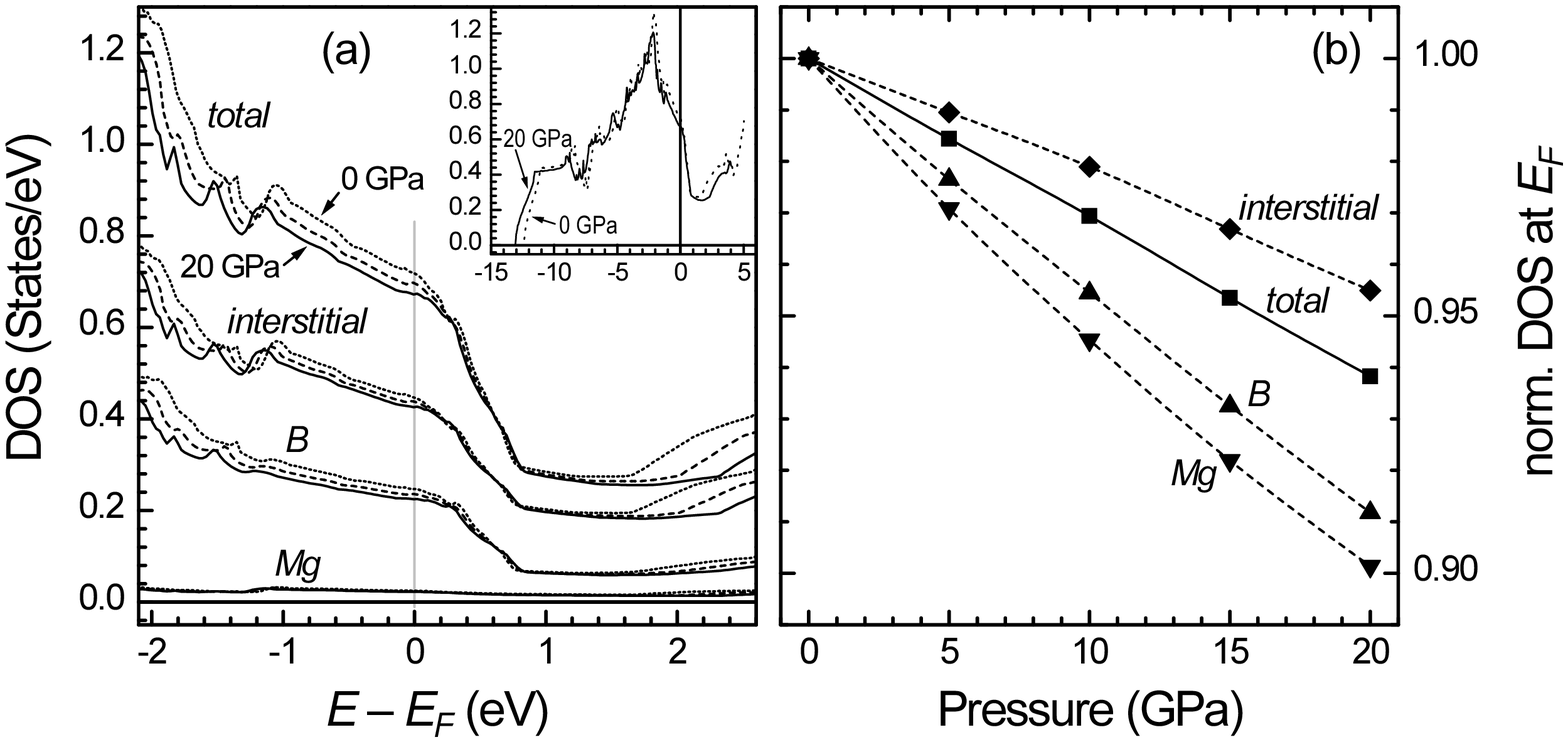}
     \vspace{3mm}
     \caption{(a) Total and partial electronic densities of states near the Fermi
     level of \MB\ at pressures of 0~GPa, 10~GPa ($V/V_0 = 0.939$), and 20~GPa
     ($V/V_0 = 0.892$) (dotted, dashed and solid lines, respectively). The inset
     shows the total DOS over a wider energy range. (b) Total and partial
     densities of state at the Fermi level as a function of pressure, normalized
     to their respective zero-pressure values.}
     \label{fig:DOS}
\end{figure}

In order to estimate the pressure dependence of \Tc, we have calculated the
electronic density of states for pressures up to 20~GPa. Figure~\ref{fig:DOS}(a)
shows the total DOS and the decomposition into partial densities near the Fermi
energy $E_F$. Qualitatively, the main effect of pressure is to spread the DOS
over an increasing range of energies. The total density of states $N(0)$ at the
Fermi level decreases continuously under pressure [Fig.~\ref{fig:DOS}(b)]. This
reduction is linear in pressure at a rate of $\d \ln N(0)/\d P = -0.31$\%/GPa.
The decomposition into partial densities crucially depends on the choice of
atomic spheres. The apparent redistribution of B and Mg partial densities into
the interstitial DOS probably results from the slight decrease of the $c/a$ ratio
and the characteristics of the charge distribution of \MB\
\cite{BSA01p,KMBA01p,IM00}.

The pressure dependence of the frequency $\omega$ of a quasi-harmonic phonon can
be expressed via the mode Gr\"{u}neisen parameter $\gamma$ as
\begin{equation}\label{eq:om1}
\frac{\d \ln \omega}{\d P} = \gamma/B_0 \quad.
\end{equation}
For solids with isotropic compression, the mode Gr\"{u}neisen parameters of
zone-center phonons are typically close to 1.  Kong
\textit{et al.}\ as well as An and Picket concluded that in
\MB\ only the branch of the in-plane $E_{2g}$ phonon of B ions exhibits a large
electron-phonon coupling \cite{KDJA01p,AP01p}. The mode Gr\"{u}neisen parameter of
the corresponding phonon of graphite was experimentally determined as
$\bar{\gamma} = 1.06$ where the anisotropic compressibility of graphite was taken
into account \cite{HBS89}. We therefore assume that in \MB\ phonons with a
frequency $\omega$ close to that of the $E_{2g}$ mode ($\omlog \approx 500$~\WN
\cite{KMBA01p,KDJA01p}) are responsible for the superconductivity and have
$\gamma=1$. Using the calculated $B_0$ of \MB\ we obtain the estimate $\d
\ln \omega/\d P \approx 1/B_0 = +0.71$\%/GPa

We now assume the Coulomb pseudopotential $\mu^*$ and the electron-ion matrix
element $I$ to be constant in first approximation. With the calculated pressure
dependences of the density of states at the Fermi level and the phonon frequency
we estimate the pressure dependence of the electron-phonon coupling constant $\d
\ln \lambda/\d P \approx -1.7$\%/GPa. Expressing the pressure dependence of \Tc\
as
\begin{equation}\label{eq:dTc}
  \frac{\d \ln \Tc}{\d P} = \frac{\partial \ln \Tc}{\partial \lambda} \frac{\d \lambda}{\d P}+
                  \frac{\partial \ln \Tc}{\partial \omlog} \frac{\d \omlog}{\d P}
\end{equation}
and using  Eqs.~(\ref{eq:McMillan}) and (\ref{eq:om1}) gives
\begin{align}
  \left.\frac{\d \ln \Tc}{\d P}\right|_{P=0} &=
  \underbrace{\frac{1.02\, \lambda}{(\lambda-\mu^*-\mu^*\lambda)^2}}_{\equiv \,\alpha}
  \cdot \frac{\d \ln \lambda}{\d P} \,
  + \left.\frac{\d \ln \omlog}{\d P} \; \right|_{P=0} \\
   &= \left. \alpha \left[\frac{\d \ln N(0)}{\d P} - \frac{\d \ln \omsquare}{\d
   P}\right]  + \frac{\d \ln \omlog}{\d P} \; \right|_{P=0} \\
   &\approx \left. \alpha \left[\frac{\d \ln N(0)}{\d P} - 2\, \frac{\d \ln \omega}{\d
   P}\right]  + \frac{\d \ln \omega}{\d P} \; \right|_{P=0} \\
   &\approx \left. \alpha \frac{\d \ln N(0)}{\d P} + (1- 2\alpha) \frac{\gamma}{B_0} \;
   \right|_{P=0} \label{eq:dTc2d}
  \quad .
\end{align}
The first term of the r.h.s.\ of Eq. (\ref{eq:dTc2d}) expresses the effect of a
change of $N(0)$ on \Tc, the second one that of the change of the phonon
frequency. For any combination of the parameters $\lambda$ and $\mu^*$ in the
relevant ranges $\lambda = 0.1-2.0$ and $\mu^* = 0-0.2$ an \emph{increasing}
phonon frequency alone would result in a \emph{decrease} of \Tc\ ($\alpha >
1/2$). The electron-phonon coupling constant of \MB\ at ambient pressure was
calculated in Ref.~\cite{KDJA01p} as $\lambda = 0.65$ and estimated in
Ref.~\cite{KMBA01p} as $\lambda \approx 0.7$. With $\lambda=0.7$ and a typical
value of $\mu^* = 0.1$ we obtain $\d \ln \Tc/\d P \approx -3.6$\%/GPa. However,
this set of parameters and $\omlog \approx 500$~\WN\ \cite{KMBA01p,KDJA01p}
underestimates $\Tc = 23$~K. If we choose an exceptionally small $\mu^* = 0.04$
in order to reproduce the measured \Tc\ we obtain a smaller $\d \ln \Tc/\d P
\approx -2.3$\%/GPa. Both of these values fall in the range marked by the
experimental results of $-4.1$\%/GPa \cite{LMC01p} and $-1.8$\%/GPa
\cite{MNRR01p}. Under the assumptions made, i.e., that the pressure dependences
of $I$ and $\mu^*$ are negligible, it is not sufficient to consider only the
change of density of states $N(0)$. It would reduce the estimated pressure
dependences of \Tc\ by a factor of $\sim$4 making them at least a factor of 2
smaller than the smallest experimental value. \emph{Thus, the increase of the
characteristic phonon frequency appears to be the dominant contribution to the
decrease of \Tc\ under pressure.}

The implications of our calculations are threefold. Firstly, the results
regarding the elastic properties show that the combination of covalent bonding
within the B layers and metallic bonding between the Mg and B sheets is quite
well-balanced in terms of compressibility. Consequently, the hydrostatic
compression is rather isotropic despite the layered crystal structure of \MB.
Secondly, Kortus \textit{et al.}\ concluded that the high critical temperature of
\MB\ can be explained in the framework of BCS theory \cite{KMBA01p}. The
agreement of our estimate of the pressure dependence of \Tc\ with experimental
results supports this view. Thirdly, our results suggest that it is the
combination of increasing phonon frequency and decreasing electronic density of
states at the Fermi level which leads to the observed decrease of the critical
temperature under pressure. Here, the phonon contribution appears to be the
dominant one.

\clearpage

% ----------------------------------------------------------------
%\bibliographystyle{prstynoetal}
%\bibliography{papers,misc,MgB2calc}

\clearpage

\end{document}